# Rotational hysteresis of the exchange anisotropy direction

# in Co/FeMn thin films

Justin Olamit and Kai Liu[*]

Physics Department, University of California,  Davis, CA 95616

(11/10/2006)

# Abstract

We have investigated the effects of rotating an applied field on the exchange anisotropy in Co/FeMn thin films.  When the applied field is initially along the cooling field direction, the longitudinal hysteresis loop has a maximum coercivity and the transverse hysteresis loop is flat, indicating that the exchange field is along the cooling field direction.  When the applied field angle is rotated away and then restored to the original field cooling direction, the exchange anisotropy direction has changed.  The rotation of the exchange field direction trails the applied field and is hysteretic.  The rotational hysteresis of the exchange field direction is due to the weak anisotropy in thin FeMn layers, and decreases with increasing FeMn thickness.

Exchange bias in ferromagnet/antiferromagnet (FM/AF) bilayers has been studied for over 50 years because of the importance of the applications in spin-valve devices and the interesting fundamental physics involved.[1-4] Exchange bias usually manifests itself as a shift of the hysteresis loop and an increase in the coercivity. Some FM/AF systems also exhibit the training effect, where the exchange bias and coercivity decrease upon repeated cycling of the hysteresis loop.[1,5] This effect shows that the AF spin structure, which is responsible for the loop pinning and coercivity change, is not static. Urazdhin and Chien showed that a rotating field could also change the AF spin structure, dragging the spins away from their initial field cooled orientation.[6] From their measurements, they could distinguish between the FM/AF interactions that lead to pinning and those that change the coercivity. Brems *et al*. showed that applying fields perpendicular to the cooling field direction could counteract the training effects, even restoring the highly asymmetrical untrained hysteresis loop often found in trainable exchange biased samples.[7]

In this paper, we study the effects of rotating the applied field on the exchange field direction of Co/FeMn thin films. When the applied field is initially along the cooling field direction, the longitudinal loop has a maximum coercivity and the transverse loop is flat. Both loops are consistent with having the exchange field along the cooling field direction.[8] As the applied field angle is changed and then restored to the original field cooling direction, the coercivity of the longitudinal loop decreases and the transverse loop develops peaks, indicating that the exchange field no longer points along the cooling field direction. To restore the longitudinal loop coercivity and transverse loop flatness, the applied field must be rotated further to the new direction along which the exchange field points. The amount of angular change in the direction of the exchange field is dependent on the thickness of the FeMn layer.

The samples were made by magnetron sputtering using solid targets in a chamber with a base pressure of ~5 x $10^{-8}$ Torr. The samples were deposited at ambient temperature on naturally oxidized Si (100) substrates with the structure of Si / Cu (200 Å) / FeMn (t) / Co (100 Å) / Cu (200 Å) cap. X-ray diffraction showed that the Cu and FeMn had the (111) texture commonly found in FM/FeMn exchange biased systems. The samples were mounted in a Princeton Measurement Corporation vibrating sample magnetometer equipped with a He gas-flow furnace and vector coils that were sensitive to moments parallel and perpendicular to the in-plane applied field. The magnetometer allowed rotations of the sample with respect to the field. To establish exchange bias, all samples were heated to 420 K, saturated in a 1000 Oe field, and then field cooled to room temperature. Immediately after field cooling, 10 hysteresis loops were measured to train out the samples, and usually training became insignificant after 5 or 6 field cycles. Longitudinal and transverse loops were measured along the cooling field direction (defined as $\theta = 0°$). The sample was then rotated counterclockwise (CCW) to a new $\theta$ (positive) and hysteresis loops were measured. Once $\theta$ was 15° away from the cooling field direction, the sample was rotated clockwise (CW) to a new $\theta$. Additional rotations in the CCW and CW directions were made to confirm the angular location of the exchange field direction (more below). For all samples, the maximum $\theta$ was 15° beyond the angles at which the transverse loops became flat.

Hysteresis loops of a sample with 60 Å of FeMn are shown in Fig. 1. The hysteresis loop measured right after the initial 10 training loops is shown in Fig. 1(a), where the applied field is parallel to the cooling field direction. The longitudinal loop has a coercivity of $H_C = 55.4$ Oe and the transverse loop is featureless. The flatness of the transverse loop indicates that there are no *net* rotations of Co moments in a direction perpendicular to the applied field during the magnetization reversal.[8] This confirms that the applied field is parallel to the exchange bias direction. The sample is then rotated CCW 15° from the cooling field direction, resulting in a longitudinal loop with a smaller coercivity ($H_C = 31.2$ Oe) and a transverse loop with large upward-pointing peaks shown in Fig. 1(b). This transverse loop is consistent with reversals by rotation toward the direction of the exchange field. Subsequently the sample is rotated back to the cooling field direction. However, the resultant loop [Fig. 1(c)] is different from that in Fig. 1(a): the longitudinal loop coercivity ($H_C = 45.3$ Oe) is smaller and the transverse loop is no longer flat. The upward-pointing transverse loop peaks suggest that the exchange field still has a component pointing upward – not along the original cooling field direction. In fact, the sample must be rotated a further 5° CW, as shown in Fig. 1(d), to achieve the longitudinal ($H_C = 56.0$ Oe) and transverse loops that most resemble those in Fig. 1(a).

The angular dependence of the transverse loop peak height, $m_{Peak}^T$, is shown in Fig. 2(a). The decreasing- and increasing-field branches of the transverse loop exhibit slight asymmetry, similar to those reported earlier.[9-12] However, the asymmetry is not severe, thus for clarity only the average peak height is plotted. The evolution of the peaks is hysteretic – $m_{Peak}^T$ emerges (0 memu at $\theta = 0°$) as $\theta$ increases (CCW rotation of sample). After $\theta = 15°$, the sample is rotated CW (towards decreasing $\theta$) and $m_{Peak}^T$ gradually decreases, but only begins to change sign at $\theta = -5°$. After $\theta = -20°$, the sample is rotated CCW again (toward increasing $\theta$) and $m_{Peak}^T$ changes direction after $\theta = +5°$. The width of this hysteresis loop ($\theta$ vs. $m_{Peak}^T$) shows that the direction of the exchange field can vary by 10° in this sample due to the change of the angle of the applied field with respect to the sample. The corresponding angular dependence of the coercivity and exchange bias of the longitudinal loop are shown in Fig. 2(b). The coercivity plot has clear peaks at ±5°, matching the pattern found with the transverse loops. This also indicates that the applied field is parallel to the exchange field direction at $\theta = \pm5°$. The exchange bias pattern is less clear. There is a well defined local minimum at +5°, but not at -5°. Other angular dependence studies have shown that the exchange bias magnitude does not always vary monotonically with angle.[9, 13, 14] Thus a simple angular trend in the exchange field was not expected.

The magnitude of the rotational hysteresis of the exchange field direction depends on the thickness, $t$, of the FeMn layer, as shown in Fig. 3. The rotational hysteresis quickly decreases with increasing FeMn thickness, and follows roughly a $1/(a + bt)$ dependence, where $a$ and $b$ are constants. The rotational hysteresis of the exchange field direction is an indication that the AF moments that are responsible for the exchange bias effect are not completely stationery. During field rotations, the AF spins were re-orientated, leading to new exchange anisotropy directions. The changing exchange anisotropy direction manifests itself as a rotational hysteresis seen here.

The rotational hysteresis increases with decreasing AF thickness as a result of the diminishing AF anisotropy. Many studies have shown that exchange bias is highly dependent on the AF layer thickness and anisotropy value.[15, 16] Below a certain anisotropy energy, which is proportional to the AF volume, the AF layer cannot pin the FM layer to create a unidirectional anisotropy, although it may contribute to a uniaxial anisotropy that increases the FM coercivity.[6]  The amount of AF anisotropy required for exchange bias has also been related to the energy required to form an AF domain wall parallel to the FM/AF interface.[17, 18]  It has been shown in FeMn that a 90 Å thick layer is needed to form a 180º domain wall.[18]  This is roughly the thickness below which most of the rotational hysteresis is observed in the present system.  During a reversal from positive to negative saturation, the FM layer applies a torque on the AF spins, causing some of them, particularly the interfacial ones, to rotate. This forms the winding structure of a (partial) domain wall.  However, some of the rotations of the AF spins may occur irreversibly, especially when the AF anisotropy is weak in a thin AF layer.  Hence, after the FM layer reverses back to positive saturation and unwinds the AF domain wall, the net AF anisotropy axis has moved to a new angle.  The irreversibility of the AF spin rotations may be explained by the existence of multiple AF anisotropy axes in a single grain where AF spins are dragged to a new anisotropy direction by the FM layer during reversals, as proposed by Hoffman to explain the training effect.[19] Indeed, the observed rotational hysteresis has similarities with the training effect as the hysteresis loop evolves with measurement history. However, in most training effect studies, the sample geometry is not changed with respect to the applied field and repeated field cycling leads to a reduction of the exchange bias. The bias field magnitude is not completely restored even after measuring loops perpendicular to the cooling field direction.[7] Here, the exchange anisotropy direction can be restored by winding the AF domain wall in the opposite direction through sample rotation. [It is important to keep the "direction" description. We don't make any promises about the magnitude of the anisotropy]

In summary, we have studied the effects of rotating the applied field with respect to the cooling field direction on Co/FeMn thin films. When the applied field is initially along the cooling field direction, the longitudinal loop has a maximum coercivity and the transverse loop is flat, indicating that the exchange field is along the cooling field direction. As the applied field angle is changed and then restored to the original field cooling direction, the coercivity of the longitudinal loop decreases and the transverse loop develops peaks, indicating that the exchange field no longer points along the cooling field direction. The angular change in the direction of the exchange field is hysteretic and more apparent in films with a thinner FeMn layer. This rotational hysteresis is due to the weaker anisotropy in the thinner FeMn layer, which leads to irreversible rotations in the FeMn layer.

This work has been supported by the ACS-PRF (#43637-AC10) and the Alfred P. Sloan Foundation.

# Figure Captions

Fig. 1.    Longitudinal loops (filled squares) and transverse loops (open circles) for FeMn (60 Å) / Co (100 Å) when the applied field is along $\theta$ = (a) 0°, (b) 15°, (c) 0° again, and (d) -5°.  The exchange field direction has rotated away from 0° after the field rotation.

Fig. 2    Angular dependence of (a) the transverse peak height, $m_{Peak}^{T}$ and (b) coercive and exchange fields for a FeMn (60 Å) / Co (100 Å) sample.  The transverse peaks vanish at $\theta$ = ±5°, coinciding with the maximum coercivity in (b) and indicating a rotational hysteresis of 10° (dashed lines).

Fig. 3    FeMn film thickness dependence of the rotational hysteresis.  The line is a fit to a modified inverse thickness dependence described in the text.

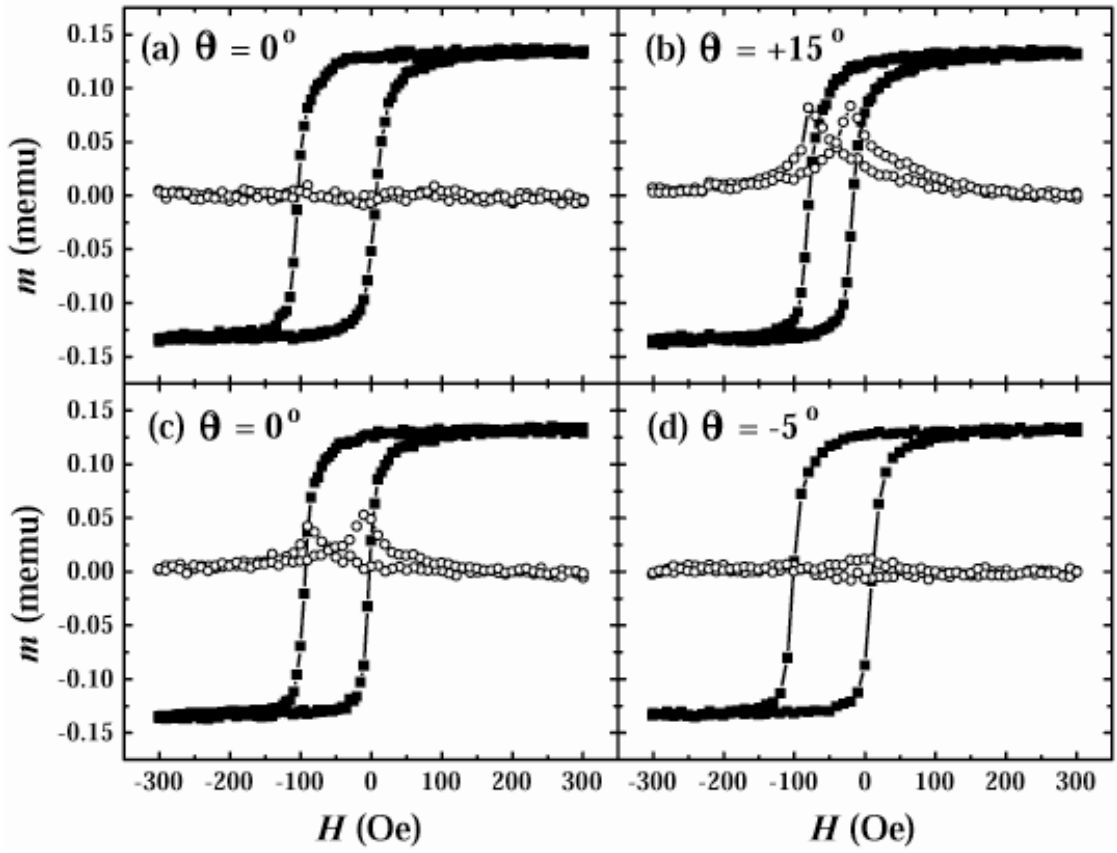

**Fig. 1**

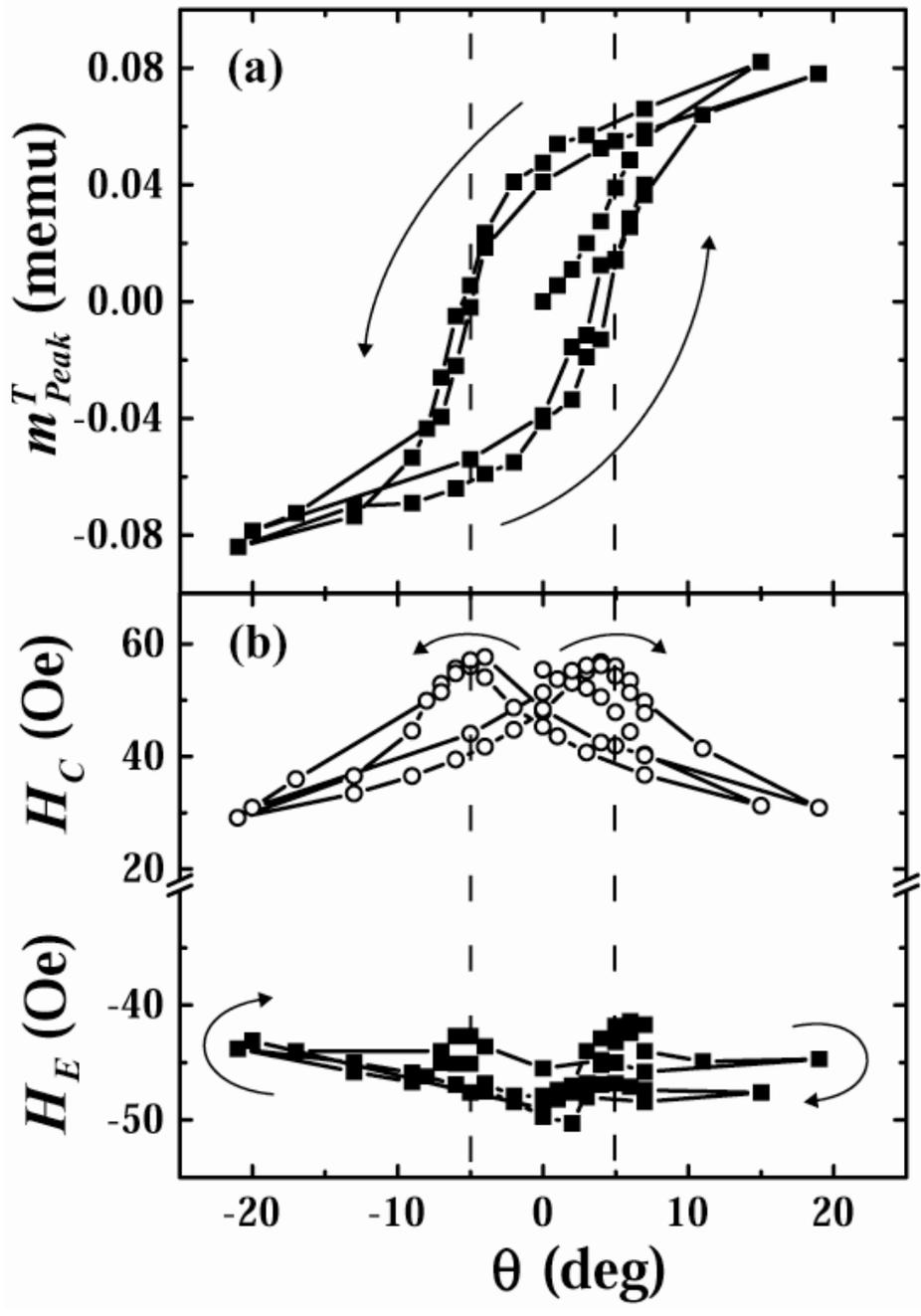

**Fig. 2**

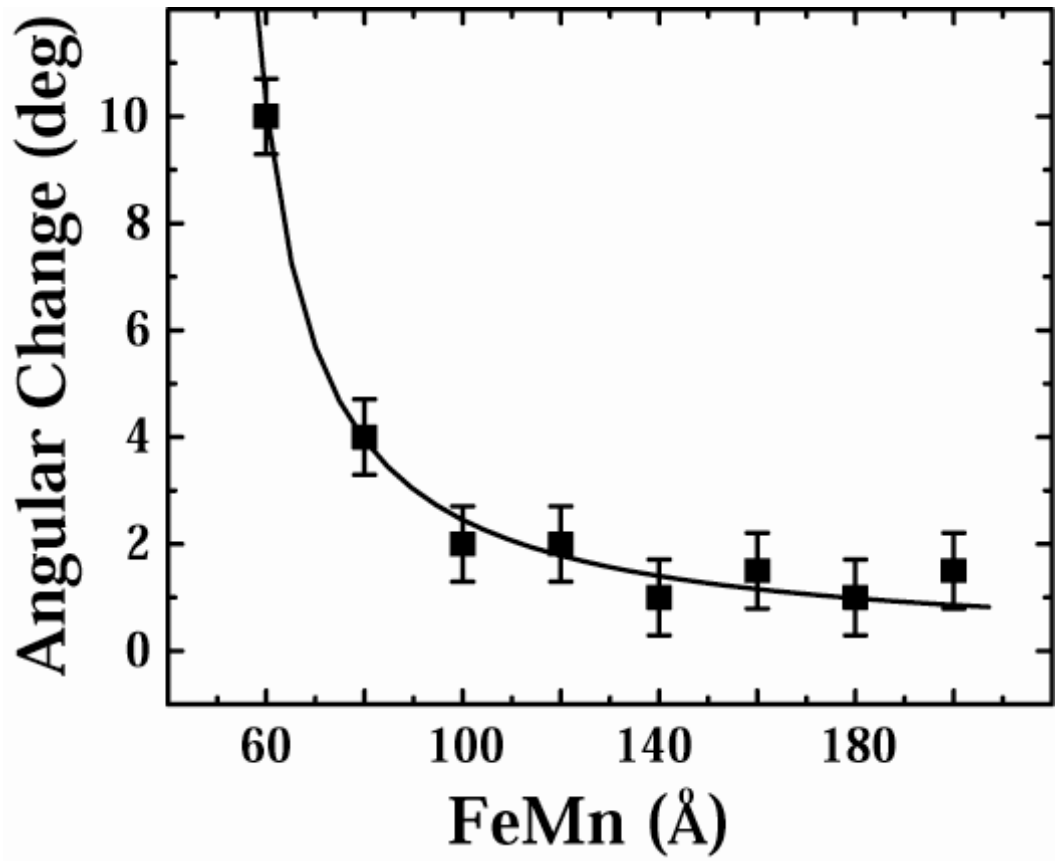

**Fig. 3**